\documentclass[apl,twocolumn,showpacs]{revtex4}
\usepackage{CJK,CJKnumb,CJKulem}
\usepackage{amssymb}
\usepackage{amsmath}
\usepackage{graphicx}
\usepackage{dcolumn}
\usepackage{bm}

\begin{document}


\title{Efficient Raman Frequency Conversion by Feedbacks of Pump and Stokes Fields}

\author{Bing Chen$^1$, Kai Zhang$^1$, Chun-Hua Yuan$^1$, Chengling Bian$^1$, \\ Cheng Qiu$^1$, L. Q. Chen$^{1,a)}$ \footnotetext{a) Electronic mail: lqchen@phy.ecnu.edu.cn.}, Z. Y. Ou$^{1,2}$ , and Weiping Zhang$^1$ }
\affiliation{$^{1}$Quantum Institute for Light and Atoms, State Key Laboratory of Precision Spectroscopy, Department of Physics,
East China Normal University, Shanghai 200062, P. R. China\\
$^{2}$Department of Physics, Indiana University-Purdue University
Indianapolis, 402 North Blackford Street, Indianapolis, Indiana 46202, USA}

\date{\today}

\begin{abstract}
 We experimentally demonstrate efficient Raman conversion to respective Stokes and anti-Stokes fields in both pulsed and continuous modes with a Rb-87 atomic vapor cell.  The conversion efficiency is about 40-50\% for the Stokes field and 20-30\% for the anti-Stokes field, respectively. This conversion process is realized with feedback of both the Raman pump and the frequency-converted fields (Stokes or anti-Stokes). The experimental setup is very simple and can be applied easily to produce the light source with larger frequency difference using other Raman media. They may have wide applications in nonlinear optics, atomic physics, quantum optics and precise measurement.

\end{abstract}


\maketitle

Efficient nonlinear interaction and frequency conversion at low light intensity is of interest in many areas of nonlinear and quantum optics because of its potential applications to high-precision spectroscopy and quantum information processing and storage. However, efficient conversion is almost always required to have high-power pumping because nonlinear coefficients are usually small in a nonlinear medium. In particular for Raman scattering, the conversion efficiency from the Raman pump field to the Stokes field is quite low. Traditionally, people could increase the conversion efficient by high-finesse optical cavity \cite{kuh,bra,mat} or stimulated Raman process \cite{ray,mis}. But these methods are complicated to some degree.

On the other hand, in the past two decades, it was discovered that nonlinear conversion can be greatly enhanced in coherent atomic ensembles. One approach is to prepare atomic spin wave before the Raman conversion process, the atomic spin wave acts as a seed to the Raman amplification process for enhanced Raman conversion. Jain et al \cite{jai} and Merriam et al \cite{mer} achieved high frequency conversion efficiencies with the help of an atomic coherence prepared via electromagnetically induced transparency \cite{boi,harris}. The conversion efficiency has reached near 40\% when the Raman write lasers have an intensity as high as several MW/cm$^2$. Recently, we demonstrated a high Raman conversion of 40\% with a low pump field intensity of 0.1 W/cm$^2$. This is achieved by first preparing a spatially distributed atomic spin wave in Rb-87 vapor with another Raman laser \cite{chen,yuan}. Nonlinear conversion efficiency can be enhanced with coherent medium prepared by counter-propagating fields and efficient intrinsic feedback \cite{fle,zib,berre,berre2}. Zibrov et al \cite{zib} observed a 4\% conversion efficiency with laser power of 300 $\mu$W and a spot size of 0.3mm. However, these schemes need other fields to prepare the atomic spin waves and can only operate in pulse mode.

In this paper, we experimentally demonstrate a simple and efficient Raman conversion scheme with coherent feedback. After the first Raman process in the forward direction, we reflect back  both the original Raman pump and the forward generated fields to the atomic medium. We obtain a conversion efficiency as high as 50\% for the Stokes field and 30\% for the anti-Stokes field with pump field power as low as 200$\mu$W. The scheme works in both pulsed and continuous wave (CW) modes with the same conversion efficiency. By beating two Stokes fields generated from a common write field, we observe a narrow line width of 10 kHz, which is determined by the decoherence time of the atomic spin wave in the medium.

\begin{figure}
\includegraphics[angle=-90,bb=78 102 406 684,scale= 0.40]{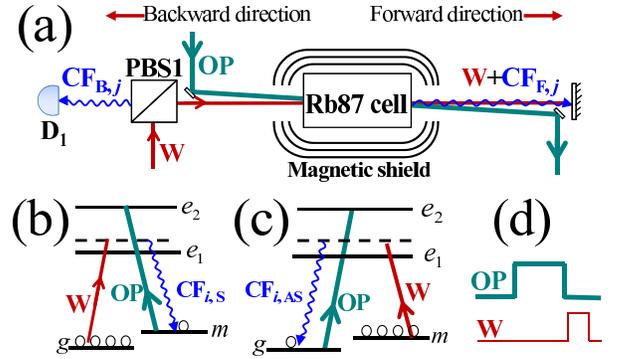}
\caption{(color online) (a) experimental arrangement. PBS: polarization beam splitter; $CF_{i, j}$: frequency-converted fields,  $i=F, B$ ($F$: forward direction; $B$: backward direction), and $j=S, AS$ ($S$: Stokes field; $AS$: anti-Stokes field); OP: optical pumping laser. (b) and (c) Energy levels of $^{87}$Rb for Stokes generation (b) and anti-Stokes generation (c); $g$ and $m$ states are the hyperfine split ground states $|5^2S_{1/2}, F=1,2\rangle$; $|e_1\rangle = |5^2P_{1/2}, F=2\rangle$ and $|e_2\rangle = |5^2P_{3/2}, F=3\rangle$. (d) Timing sequence.}
\end{figure}

The experimental setup is shown in the Fig.1(a). The protocol is based on the feedback Raman process in a pure $^{87}$Rb atomic ensemble. The $^{87}$Rb atoms are contained in a 50mm long glass cell with paraffin coating. The cell is placed inside a four-layer $\mu$-magnetic shielding to reduce stray magnetic fields and is heated up to 70$^o$C using a bifilar resistive heater. The energy levels of $^{87}$Rb atom are given in Figs.1(b)-(c) together with laser frequencies. The lower two energy states $|g\rangle$ and $|m\rangle$ are the hyperfine split ground states $5S_{1/2}$ $(F=1,2)$ with a frequency difference of 6.87GHz and the two higher energy states $|e_1\rangle$ and $|e_2\rangle$ are the excited states ($5P_{1/2},  {5P_{3/2}}$). An optical pumping field (OP) is used to prepare the atoms in either $|g\rangle$ or $|m\rangle$ state. $W$ is the Raman pump field with a diameter of 1.0 mm. Fig.1(b) is for Stokes generation while Fig.1(c) is for anti-Stokes generation. If we tune all laser frequency ($W$ and OP) as shown in Fig.1(b), $W$ field couples the states $|e_1\rangle$ and $|g\rangle$, the frequency of the generated converted field is equal to the frequency of $W$ minus 6.87GHz, corresponding to Stokes field generation. If we tune the laser frequency as in Fig.1(c), the frequency of the converted field is equal to the frequency of $W$ plus 6.87GHz, corresponding to anti-Stokes field generation.  A mirror is placed behind the atomic cell to reflect both $W$ and the generated fields back into the atomic cell for subsequent efficient conversion. This arrangement of the mirror is a crucial part in the setup. The generated fields are separated from $W$ field by a polarization beam splitter (PBS1) because their polarizations are orthogonal to each other. Photo-detector D1 is used to measure the generated fields.

\begin{figure}
\includegraphics[angle=-90,bb=63 89 492 660,scale=0.40]{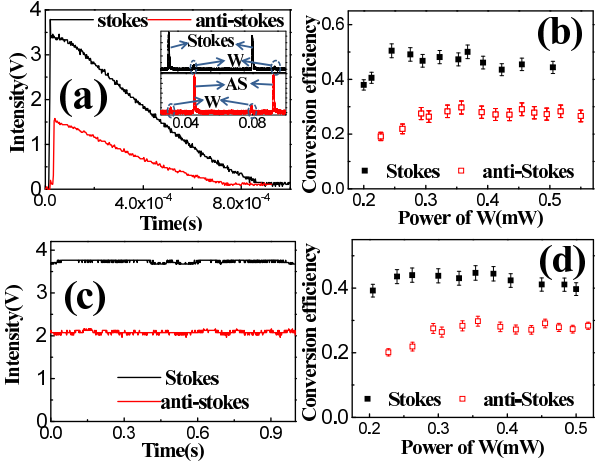}
\caption{(color online) (a) and (c) The temporal behavior of the converted field when $W$ field is in (a) pulsed mode and (c) CW mode; the inset is the frequency analysis of the converted field by a FP cavity. (b) and (d) Conversion efficiency from $W$ to the generated fields in (b) pulsed mode and (d) CW mode; black solid square is for Stokes field and red hollow square is for anti-Stokes field.}
\end{figure}

\begin{figure}
\includegraphics[angle=0,scale= 0.55]{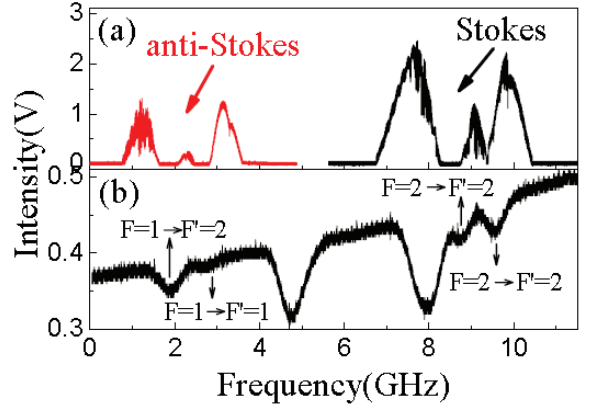}
\caption{(color online) (a) Intensity of the converted field as the frequency of $W$ field is scanned; the left red curve is for anti-Stokes and the right black curve is for Stokes. (b) Absorption spectrum of Rb (85 and 87) for frequency calibration in (a).}
\end{figure}

Firstly, we perform the experiment in pulsed mode with a timing sequence shown in Fig.1(d). W and OP lasers are chopped into pulse by acoustic-optic modulators (AOM).  The OP pulse lasts 200$\mu s$ to prepare all atoms in the ground state $|g\rangle$ or the state $|m\rangle$. Then the $W$ laser turns on for 1000 $\mu s$ and interacts with the atomic ensemble to generate the Stokes or anti-Stokes light.  The temporal behavior of the converted fields is shown in Fig. 2(a). The intensity peaks quickly and decreases with the time due to the decay of the atomic coherence. The coherence time of the paraffin cell is 500$\mu s$. The inset in Fig.2(a) is the frequency analysis of the generated field by a Fabry-Perot cavity (FP). Almost all part are the generated field with a small leaked $W$ also shown. We measure the conversion efficiency from $W$ laser to the generated field and the results are given in Fig.2(b). The efficiency ranges around 40-50\% for Stokes and 20-30\% for anti-Stokes, depending on the power of $W$.

Next, we perform the experiment in continuous wave (CW) mode by applying continuously the OP field and $W$ field. A steady frequency-converted field is generated as shown in Fig.2(c). The conversion efficiency is almost the same as the pulsed case, as shown in Fig.2(d) where we plot the efficiency as a function of the power of $W$. In the CW mode, we can check the tuning range of the generated field  by scanning the frequency of $W$. The result is shown in Fig.3 together with the absorption spectrum of $^{87}$Rb for frequency calibration. The right black and left red curves are for the Stokes and anti-Stokes fields, respectively. The red and black curves each consist of three peaks, which match well the Raman gain profile. The two large side peaks correspond to blue and red detuned Raman process, respectively. The small middle peak is due to the crossover of the two hyperfine lines of $5^2S_{1/2}, F \rightarrow 5^2P_{1/2}, F^{\prime}=1,2$ transitions. The frequency difference between $5^2P_{1/2}, F^{\prime}=1,2$ energy levels is 800MHz, while the Doppler broadening at cell temperature of 70 degree is about 600-700MHz. From this figure, we obtain a tuning range of 3.0 and 4.0GHz for anti-Stokes and Stokes, respectively.

In the CW mode, we are able to look at the coherence property of the converted field. To do this, we split $W$ into two beams and convert each beam to Stokes field. We then superimpose the two generated fields for interference. AC Stark effect leads to a slight difference between the frequencies of the two generated fields because of the difference in power and geometry in the interaction of the two beams with atoms. So we observe a beat signal shown in the inset of Fig.4. Fourier transformation of the beat signal is recorded by a spectrum analyzer and shown in Fig.4. The line width of the beat signal is about 10kHz, corresponding to a coherence time of 500 $\mu s$. This is in the same order as the decoherence time of atoms in a paraffin-coated cell.

\begin{figure}
\includegraphics[angle=-90,bb=54 450 544 126,scale= 0.35]{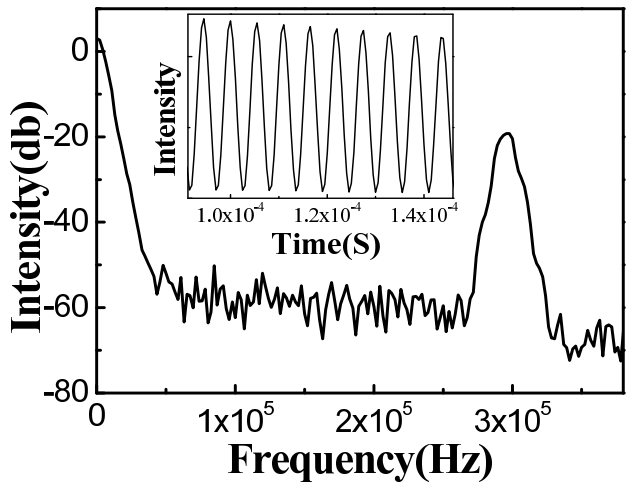}
\caption{(color online) Demonstration of coherence of the generated field: beating signal (inset) and its Fourier transformation between two similarly generated fields.}
\end{figure}

\begin{figure}
\includegraphics[angle=-90,bb=49 77 358 526,scale= 0.5]{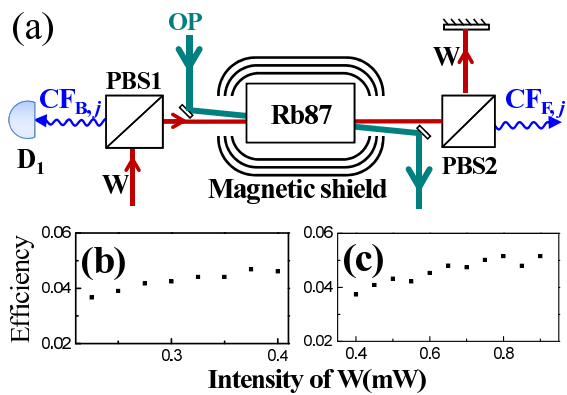}
\caption{(color online) experimental sketch of Raman conversion process without feedback. The conversion efficiency in (b) pulsed mode and (c) CW mode.}
\end{figure}
Finally, to show the enhancement effect of the feedback, we add a PBS between the flat mirror and the cell to separate the pump field $W$ and the forward generated field. We reflect back only the pump field $W$. The experimental arrangement is shown in Fig.5(a). The efficiency is given in Fig.5(b) and (c) for the pulsed and CW cases, respectively. The efficiency is around a few percent, an order of magnitude smaller than the scheme with feedback of the forward generated field. This clearly demonstrates the advantage of the scheme with feedback.

Let us understand the feedback mechanism here. When the pump field $W$ in the forward direction interacts with the atomic ensemble in the ground state by Raman scattering, a converted field in forward direction and an atomic spin wave are generated. The atomic spin wave stays in the cell. When the forward pump and the converted fields are both reflected back to the cell by the flat mirror, the Raman process by the pump field in the backward direction will be stimulated by the reflected forward converted field and enhanced by the previously produced atomic spin wave at the same time. An interference effect occurs between backward converted fields produced by the two mechanisms because of the phases correlation between the forward converted field and the atomic spin wave \cite{mis,bian}. The co-propagation of the reflected fields will lead to in-phase constructive interference and thus enhanced conversion efficiency.

In conclusion, we have demonstrated an efficient way to make Raman conversion with feed-back. The conversion efficiency is about 40\% for Stokes field and 20\% for anti-Stokes field with as little as a few hundreds of $\mu$W of Raman pump. Such a scheme can replace traditional techniques such as EOM and AOM to obtain a large frequency shift for studying light interaction with atoms such as the EIT effect for manipulation of atomic spin waves \cite{harris} and Raman atomic interferometer \cite{chu}.

This work was supported by the National Basic Research Program of China (973 Program
grant no. 2011CB921604), the National Natural Science Foundation of China (grant numbers
11004058, 11004059, 11129402, J1030309, 11274118, 10828408 and 10588402) and the Program of Shanghai Subject Chief Scientist (grant number 08XD14017).

\end{document}